\RequirePackage{lineno}
\documentclass[aps,prl,twocolumn,superscriptaddress]{revtex4}  
\usepackage{graphicx}
\usepackage{dcolumn}
\usepackage{bm}
\usepackage{amsmath}
\usepackage{url}
\usepackage{float}
\usepackage{hhline}
\usepackage{comment}
\usepackage{xcolor}

\begin{document}

\title{First measurement of quasi-elastic $\Lambda$ baryon production in muon anti-neutrino interactions in the MicroBooNE detector}

\newcommand{\ANL}{Argonne National Laboratory (ANL), Lemont, IL, 60439, USA}
\newcommand{\Bern}{Universit{\"a}t Bern, Bern CH-3012, Switzerland}
\newcommand{\BNL}{Brookhaven National Laboratory (BNL), Upton, NY, 11973, USA}
\newcommand{\UCSB}{University of California, Santa Barbara, CA, 93106, USA}
\newcommand{\Cambridge}{University of Cambridge, Cambridge CB3 0HE, United Kingdom}
\newcommand{\CIEMAT}{Centro de Investigaciones Energ\'{e}ticas, Medioambientales y Tecnol\'{o}gicas (CIEMAT), Madrid E-28040, Spain}
\newcommand{\Chicago}{University of Chicago, Chicago, IL, 60637, USA}
\newcommand{\Cincinnati}{University of Cincinnati, Cincinnati, OH, 45221, USA}
\newcommand{\CSU}{Colorado State University, Fort Collins, CO, 80523, USA}
\newcommand{\Columbia}{Columbia University, New York, NY, 10027, USA}
\newcommand{\Edinburgh}{University of Edinburgh, Edinburgh EH9 3FD, United Kingdom}
\newcommand{\FNAL}{Fermi National Accelerator Laboratory (FNAL), Batavia, IL 60510, USA}
\newcommand{\Granada}{Universidad de Granada, Granada E-18071, Spain}
\newcommand{\Harvard}{Harvard University, Cambridge, MA 02138, USA}
\newcommand{\IIT}{Illinois Institute of Technology (IIT), Chicago, IL 60616, USA}
\newcommand{\KSU}{Kansas State University (KSU), Manhattan, KS, 66506, USA}
\newcommand{\Lancaster}{Lancaster University, Lancaster LA1 4YW, United Kingdom}
\newcommand{\LANL}{Los Alamos National Laboratory (LANL), Los Alamos, NM, 87545, USA}
\newcommand{\Louisiana}{Louisiana State University, Baton Rouge, LA, 70803, USA}
\newcommand{\Manchester}{The University of Manchester, Manchester M13 9PL, United Kingdom}
\newcommand{\MIT}{Massachusetts Institute of Technology (MIT), Cambridge, MA, 02139, USA}
\newcommand{\Michigan}{University of Michigan, Ann Arbor, MI, 48109, USA}
\newcommand{\Minnesota}{University of Minnesota, Minneapolis, MN, 55455, USA}
\newcommand{\NMSU}{New Mexico State University (NMSU), Las Cruces, NM, 88003, USA}
\newcommand{\Oxford}{University of Oxford, Oxford OX1 3RH, United Kingdom}
\newcommand{\Pitt}{University of Pittsburgh, Pittsburgh, PA, 15260, USA}
\newcommand{\Rutgers}{Rutgers University, Piscataway, NJ, 08854, USA}
\newcommand{\SLAC}{SLAC National Accelerator Laboratory, Menlo Park, CA, 94025, USA}
\newcommand{\SDSMT}{South Dakota School of Mines and Technology (SDSMT), Rapid City, SD, 57701, USA}
\newcommand{\Maine}{University of Southern Maine, Portland, ME, 04104, USA}
\newcommand{\Syracuse}{Syracuse University, Syracuse, NY, 13244, USA}
\newcommand{\TelAviv}{Tel Aviv University, Tel Aviv, Israel, 69978}
\newcommand{\Tennessee}{University of Tennessee, Knoxville, TN, 37996, USA}
\newcommand{\UTA}{University of Texas, Arlington, TX, 76019, USA}
\newcommand{\Tufts}{Tufts University, Medford, MA, 02155, USA}
\newcommand{\VTech}{Center for Neutrino Physics, Virginia Tech, Blacksburg, VA, 24061, USA}
\newcommand{\Warwick}{University of Warwick, Coventry CV4 7AL, United Kingdom}
\newcommand{\Yale}{Wright Laboratory, Department of Physics, Yale University, New Haven, CT, 06520, USA}

\affiliation{\ANL}
\affiliation{\Bern}
\affiliation{\BNL}
\affiliation{\UCSB}
\affiliation{\Cambridge}
\affiliation{\CIEMAT}
\affiliation{\Chicago}
\affiliation{\Cincinnati}
\affiliation{\CSU}
\affiliation{\Columbia}
\affiliation{\Edinburgh}
\affiliation{\FNAL}
\affiliation{\Granada}
\affiliation{\Harvard}
\affiliation{\IIT}
\affiliation{\KSU}
\affiliation{\Lancaster}
\affiliation{\LANL}
\affiliation{\Louisiana}
\affiliation{\Manchester}
\affiliation{\MIT}
\affiliation{\Michigan}
\affiliation{\Minnesota}
\affiliation{\NMSU}
\affiliation{\Oxford}
\affiliation{\Pitt}
\affiliation{\Rutgers}
\affiliation{\SLAC}
\affiliation{\SDSMT}
\affiliation{\Maine}
\affiliation{\Syracuse}
\affiliation{\TelAviv}
\affiliation{\Tennessee}
\affiliation{\UTA}
\affiliation{\Tufts}
\affiliation{\VTech}
\affiliation{\Warwick}
\affiliation{\Yale}

\author{P.~Abratenko} \affiliation{\Tufts}
\author{D.~Andrade~Aldana} \affiliation{\IIT}
\author{J.~Anthony} \affiliation{\Cambridge}
\author{L.~Arellano} \affiliation{\Manchester}
\author{J.~Asaadi} \affiliation{\UTA}
\author{A.~Ashkenazi}\affiliation{\TelAviv}
\author{S.~Balasubramanian}\affiliation{\FNAL}
\author{B.~Baller} \affiliation{\FNAL}
\author{G.~Barr} \affiliation{\Oxford}
\author{J.~Barrow} \affiliation{\MIT}\affiliation{\TelAviv}
% ...

\collaboration{The MicroBooNE Collaboration}
\thanks{microboone\_info@fnal.gov}\noaffiliation

\begin{abstract}
We present the first measurement of the cross section of Cabibbo-suppressed $\Lambda$ baryon production, using data collected with the MicroBooNE detector when exposed to the neutrinos from the Main Injector beam at the Fermi National Accelerator Laboratory. The data analyzed correspond to $2.2 \times 10^{20}$ protons on target of neutrino mode running and $4.9 \times 10^{20}$ protons on target of anti-neutrino mode running. An automated selection is combined with hand scanning, with the former identifying five candidate $\Lambda$ production events when the signal was unblinded, consistent with the GENIE prediction of $5.3 \pm 1.1$ events. Several scanners were employed, selecting between three and five events, compared with a prediction from a blinded Monte Carlo simulation study of $3.7 \pm 1.0$ events. Restricting the phase space to only include $\Lambda$ baryons that decay above MicroBooNE's detection thresholds, we obtain a flux averaged cross section of $2.0^{+2.2}_{-1.7} \times 10^{-40}$ cm$^2/$Ar, where statistical and systematic uncertainties are combined.

\end{abstract}

\maketitle


In this letter we describe the measurement of the cross section for Cabibbo-suppressed (direct) $\Lambda$-baryon production in a restricted phase space using the MicroBooNE detector. The MicroBooNE detector~\cite{MicroBooNE:2016pwy} is a liquid argon time projection chamber (LArTPC) with several years of accumulated data using the neutrinos produced by the Main Injector (NuMI) beam~\cite{MINERvA:2016iqn,Adamson:2015dkw} at the Fermi National Accelerator Laboratory. This enables studies of rare processes such as the direct production of $\Lambda$ baryons in interactions between muon anti-neutrinos and argon in the detector:

\begin{align}
\Bar{\nu}_{\mu} + \textrm{Ar} \to \mu^+ + \Lambda + X,
\end{align}

\noindent where $X$ denotes additional final state particles with no strangeness. This process is poorly constrained by existing measurements~\cite{Ammosov:1986jn,Eichten:1972bb,Erriquez1,Erriquez2,Brunner:1989kw,Fanourakis:1980} and is sensitive to the physics of the underlying neutrino interaction and nuclear effects, including nucleon form factors and axial masses, hyperon-nucleus potentials, and final state interactions~\cite{Thorpe:2020tym,Sobczyk:2019uej,Singh:2006xp}. Such a process constitutes a potential source of background in proton decay experiments, such as DUNE~\cite{DUNE:2020fgq,DUNE:2020ypp} and Hyper-Kamiokande~\cite{Hyper-Kamiokande:2018ofw}. If the $\Lambda$ baryon undergoes a secondary interaction with a nucleon, a kaon can be produced, mimicking the $p \to K + \nu$ signal in these experiments.  Additionally, this process is exclusively the result of anti-neutrino interactions and therefore could be used to constrain contamination from anti-neutrinos in a neutrino beam.

This letter describes the measurement of a restricted phase space cross section for direct $\Lambda$ production using the MicroBooNE detector. To maximize statistics, we combine data collected when the NuMI beam was operating in its neutrino (forward horn current, FHC) and anti-neutrino (reverse horn current, RHC) modes.

\begin{figure}
    \centering
    \includegraphics[width=\linewidth]{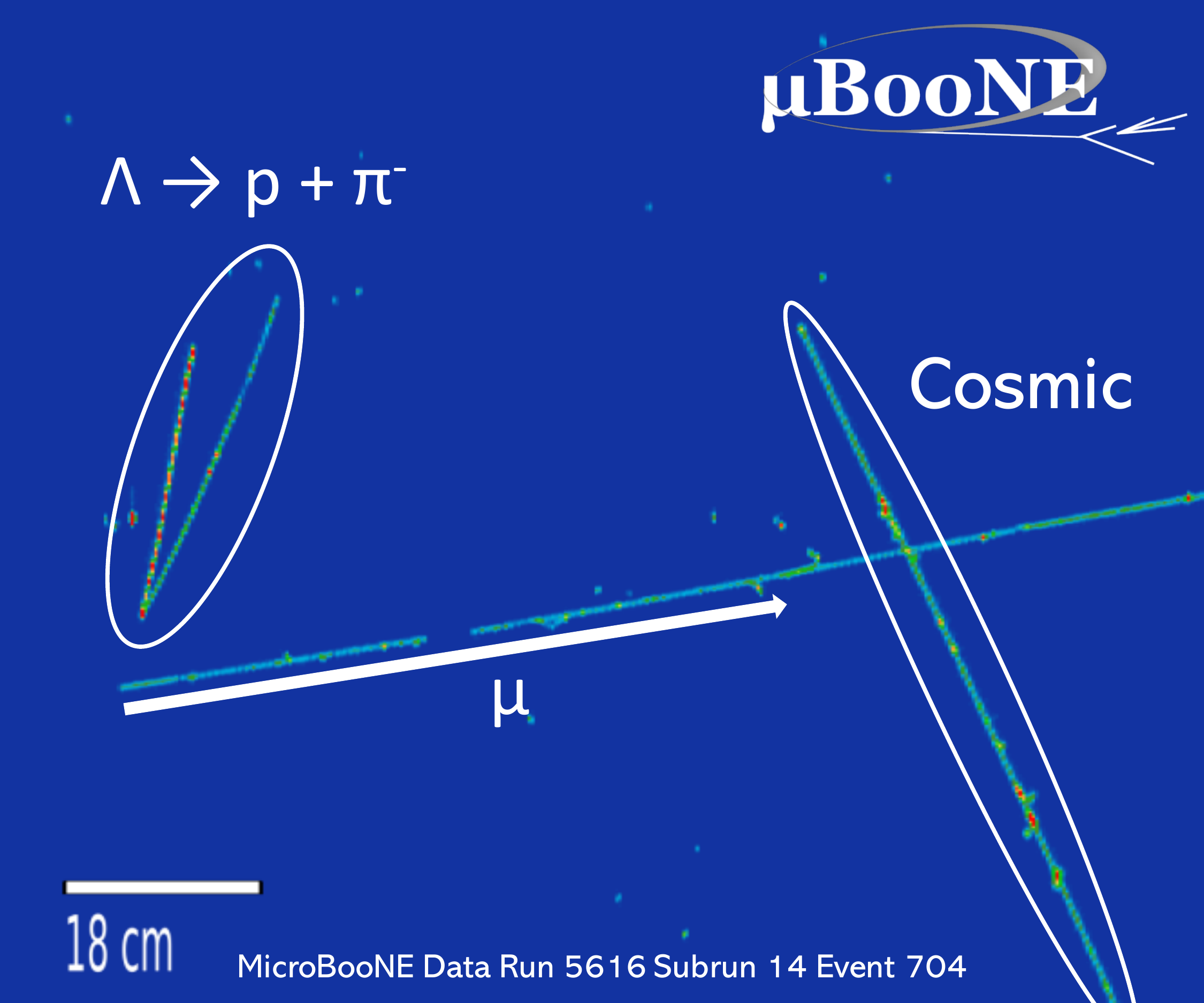}
    \caption{A candidate $\bar{\nu}_{\mu} + \textrm{Ar} \to \mu^+ + \Lambda$ interaction observed in MicroBooNE data. A cosmic ray is also reconstructed in the event. The ionization is displayed by the color scale, with green/red indicating weaker/stronger intensity. There is a region without active wires partway along the muon track.}
    \label{fig:Evd}
\end{figure}

The selection searches for muon-anti-neutrino interactions with argon nuclei, contained within the fiducial volume defined in Ref.~\cite{MicroBooNE:2021zul}, in which a $\Lambda$ is produced through the strangeness-violating quasi-elastic process and subsequently decays to a proton and negatively charged pion. This decay produces a distinctive V shaped signature in the detector, an example of which can be seen in Fig.~\ref{fig:Evd}.


We employ the GENIE~\cite{GENIE} event generator to simulate neutrino interactions inside the MicroBooNE cryostat and surrounding material, in conjunction with {\sc Geant}~4~\cite{GEANT4:2002zbu} for particle propagation and secondary interactions, followed by a simulation of the detector response to the interactions of those particles. The event selection employs the Pandora multi-algorithm reconstruction framework~\cite{MicroBooNE:2017xvs} which identifies a reconstructed neutrino vertex and the associated particles, which are classified as either tracks or showers. 

To isolate $\Lambda$ production events, we apply a number of criteria. Only data collected using a trigger synchronized with the spills of the NuMI beam are used to prevent any contamination from neutrinos produced by the Booster Neutrino Beam~\cite{MiniBooNE:2008hfu}. A neutrino vertex must be reconstructed in the fiducial volume defined in an earlier analysis~\cite{MicroBooNE:2021zul} with at least three associated tracks and no showers. Particle identification (PID) scores~\cite{MicroBooNE:2021ddy} are calculated for each track indicating whether they are muon-like or proton-like, and the longest muon-like track is selected as the muon candidate.

An array of boosted decision trees~\cite{Hocker:2007ht} is employed to generate a response score from several variables such as PID scores~\cite{MicroBooNE:2021ddy} and the Pandora track/shower classification score~\cite{VanDePontseele:2020tqz} to select a pair of tracks consistent with the $\Lambda \to p + \pi^-$ decay. The momenta of the proton and $\pi^-$ are estimated from the lengths and directions of their respective tracks; the sum of these quantities gives the momentum of the $\Lambda$ candidate. The reconstructed invariant mass $W$ and angular deviation, defined as the angle between the line connecting the primary vertex to the decay vertex and the momentum vector of the $\Lambda$ candidate, are calculated. Events with $1.09 < W < 1.19$ GeV and angular deviation $< 14^{\circ}$ are retained. 

After deconvolution and noise removal~\cite{MicroBooNE:2018swd,MicroBooNE:2018swe,MicroBooNE:2017qiu}, the charge deposited on the wires of the detector can be used to visualize the trajectories of particles produced in the interaction. This is the information displayed in Fig.~\ref{fig:Evd}, in which the green/red regions indicate nonzero activity. This is analyzed to determine if the muon candidate and $\Lambda$ candidate form separate ``islands'' of activity. This tests whether the $\Lambda$ candidate created a true secondary vertex, a feature which discriminates $\Lambda$ production from background processes with similar kinematics. The MicroBooNE detector records information from three planes of parallel wires, one of which is oriented vertically while the others are angled at $\pm 60^{\circ}$ from the vertical, providing different views of the interaction. This test is performed separately using information from each of the wire planes, enabling identification of the decay vertex even when the orientation of the event makes this difficult when viewed from one of the planes. The island finding algorithm is described in detail in the supplemental material~\cite{SupplementalMaterial}.

After the automated event selection is complete, the background primarily consists of other sources of $\Lambda$ baryons and hyperons, including other quasi-elastic-like interactions (``direct''), deep inelastic scattering (DIS), and resonant interactions (RES). A small number of background events produced by secondary interactions of neutrons is predicted, in which charged particles are liberated, most commonly $p\pi^{\pm}$ and $pp$ pairs, which can lead to a similar V shape. The remainder is due to mis-reconstruction of events. Cosmic rays and out-of-cryostat neutrino interactions (``dirt'') are included in the simulation but none pass the selection. The numbers of events from each category selected in the MC simulation are shown in Table~\ref{tab:SelectedEvents}. The final efficiency of the automated selection is $6.8\%$.

A visual scan of event displays of the selected data is performed to remove background selected due to reconstruction problems. To evaluate the background rejection power and reliability of this technique, a blinded study with five scanners was completed, using Monte Carlo (MC) simulated events that had passed the automated selection. The mean selection rates of the five scanners are calculated for the signal and each category of background, which multiplied by the number of events of each type passing the automated selection yields a new set of predictions. The signal and rates of each source of background, before and after including the visual scanning, are compared in Table~\ref{tab:SelectedEvents}. To maintain blindness in the final measurement, a separate set of MC simulated events is mixed with the data from the signal region to conceal the number of data events from the scanners. The visual scan reduces the final selection efficiency to $6.1\%$.

As an additional test to confirm the validity of the visual scanning procedure, an alternative analysis without hand scanning was performed. To constrain the predicted background due to reconstruction problems, a sideband is employed, created by inverting the cuts applied to the invariant mass and angular deviation. The result of this alternative selection and constraint are also used to calculate the cross section and yield a result consistent with the one obtained when performing the visual scan, albeit with slightly poorer sensitivity. This constraint method is described in the supplemental material~\cite{SupplementalMaterial}.

\begin{table}
    \centering
    \begin{tabular}{ccc}
    \hhline{===} 
        Event Category & Selected MC & After Visual Scan \\
        \hline
        Signal & $2.5 \pm 0.6$ & $2.3 \pm 0.6$ \\
        \hline
        Other $\Lambda$ & $0.7 \pm 0.2$ & $0.5 \pm 0.3$ \\
        Other hyperons & $1.0 \pm 0.5$ & $0.7 \pm 0.5$ \\
        Neutrons & $0.3 \pm 0.1$ & $0.1 \pm 0.1$ \\
        Mis-reconstruction & $0.9 \pm 0.4$ & $0.1 \pm 0.1$ \\
        \hline 
        Total Background & $2.8 \pm 0.9$ & $1.4 \pm 0.8$ \\
        \hhline{===}
    \end{tabular}
    \caption{Events selected from Monte Carlo (MC) simulation using standard GENIE model parameters, before and after the hand scanning selection efficiencies are applied. Combined MC simulation statistical and systematic uncertainties are shown.}
    \label{tab:SelectedEvents}
\end{table}

Two sources of flux uncertainty are considered: the hadron production modeling and the beamline geometry. The flux uncertainty in the predicted number of signal events passing the event selection using the default simulation is small (approximately 11\%) due to the high neutrino energy threshold for $\Lambda$ production. The uncertainties on the production rates for the hadrons that subsequently decay into neutrinos dominate the flux uncertainty in this energy region~\cite{AliagaSoplin:2016shs}. The uncertainty in the background due to the flux is approximately 10\%. The uncertainty in the total $\bar{\nu}_{\mu}$ flux is around 23\%.

To determine the uncertainties from the models used to derive the cross sections for background neutrino interactions, we use the results of the fits described in Ref.~\cite{MicroBooNE:2021ccs}, with 44 parameters varied in parallel to produce 600 variations. In addition, we use predictions from 8 alternative models to estimate uncertainties resulting from parameters that are difficult to vary continuously. A 100\% uncertainty is assumed for the background from other CCQE-like hyperon production processes. The overall uncertainty due to background neutrino interaction cross sections is around 35\%.

Secondary interactions in the argon outside the nuclear remnant are described by {\sc Geant} 4~\cite{GEANT4:2002zbu}; we use the Geant4Reweight~\cite{Calcutt:2021zck} package to determine the uncertainties from the description of these reinteractions by varying proton, charged pion, and $\Lambda$ baryon interaction cross sections. We assume an uncertainty of $20\%$ on the proton and $\Lambda$ interaction cross sections, while for the charged pions a pair of multi-target, multi-channel fits are performed using external data to extract uncertainties on the cross sections of individual interaction channels, as described in Ref.~\cite{Calcutt:2021zck}.  To include uncertainties on the neutron interaction cross sections, a fit is performed to data from the CAPTAIN experiment~\cite{CAPTAIN:2019fxo}, yielding an uncertainty of 26\% on the total $n$-Ar cross section. This uncertainty is included by re-scaling the rate of selected events containing secondary interactions of neutrons by $\pm 26$\%. The resulting uncertainties in the predicted signal and background are 3\% and 6\% respectively.


To assess the uncertainties on the modeling of the detector response, a set of simulated neutrino interactions in MicroBooNE is fed into several alternative detector models. These models vary the quantity of scintillation light produced, the wire response~\cite{MicroBooNE:2021roa,MicroBooNE:2019efx}, the space charge~\cite{MicroBooNE:2019koz,MicroBooNE:2020kca}, and the recombination of argon ions. The simulated detector response from each of these models is reconstructed and the $\Lambda$ selection criteria are applied. The differences between the numbers of events selected using the standard detector simulation and these alternative models are used to calculate a systematic uncertainty. The uncertainties in the signal and background due to the detector response model are 7\% and 16\% respectively.

The selection efficiencies and background acceptance rates of the five individual scanners are treated as five sets of predictions, the spread of which is used as an uncertainty. This is the largest source of uncertainty in the background prediction, contributing a fractional uncertainty of approximately 45\%, while the uncertainty in the signal due to the visual scanning is estimated to be 7\%.

Lastly, MC simulation calculations of the selection efficiency show some non-uniformity with respect to the shape $\Lambda$ baryon production cross section. MC simulation events generated with the GENIE and NuWro generators were analyzed, producing two separate estimates of the selection efficiency. The difference between these two efficiency estimates, a relative change of approximately 19\%, was adopted as another uncertainty in the selection efficiency. Variations in the model parameters, described in the supplemental material~\cite{SupplementalMaterial}, were also studied but the resulting effects on the efficiencies are smaller.

MicroBooNE is sensitive to protons and charged pions with momenta $> 0.3$~GeV/c and $>0.1$~GeV/c respectively, and the phase space of the measured cross section is therefore restricted. The relation between the restricted phase space cross section, $\sigma_R$ and the total cross section depends on the momentum distribution of the $\Lambda$ baryons produced, and is described in the supplemental material~\cite{SupplementalMaterial}. $\sigma_R$, is related to the number of events observed in data, $N_{\textrm{obs}}$, by:

\begin{align}
\sigma_R = \frac{N_{\textrm{obs}} - B}{T\Phi\Gamma\epsilon},
\end{align}

\noindent where $B$ is the predicted number of background events, $T$ the number of argon nuclei in the fiducial volume, $\Phi$ the total muon anti-neutrino flux, $\Gamma = 0.64$ the branching fraction for the process $\Lambda \to p + \pi^-$~\cite{ParticleDataGroup:2020ssz}, and $\epsilon$ the average selection efficiency. 

To account for asymmetries in the statistical uncertainties on the data and MC simulation, we employ a Bayesian procedure to calculate the full posterior distribution on the extracted cross section. Bayesian posterior distributions of the selection efficiency and background acceptance are estimated with the TEfficiency class~\cite{TEfficiency}. The data statistical uncertainty is included by applying Bayes' theorem to the Poisson likelihood function, $P(N_{\rm obs}|N)$:
\begin{equation}
P(N|N_{\rm obs}) = \frac{P(N_{\rm obs}|N)P(N)}{\int P(N_{\textrm{obs}}|N)P(N)dN},
\end{equation}
from which we sample values of $N$. $P(N)$ is the Bayesian prior of $N$; uniform priors are used for $N$, the background acceptance and selection efficiency.

\begin{table}[H]
    \centering
    \begin{tabular}{c|ccc}
    & $\epsilon$ & $\Phi$ & $B$ \\
    \hline
    $\epsilon$ & 0.04572 & -0.00116 & 0.03237 \\
    $\Phi$ & -0.00116 & 0.05339 & 0.01887 \\
    $B$ & 0.03237 & 0.01887 & 0.33123 \\
    \end{tabular}
    \caption{Fractional covariance matrix between the uncertainties on the selection efficiency $\epsilon$, the $\bar{\nu}_{\mu}$ flux $\Phi$, and the predicted number of background events $B$.}
    \label{tab:CovarianceMatrix}
\end{table}

\begin{figure*}[htb]
    \centering
    \includegraphics[width=0.7\linewidth]{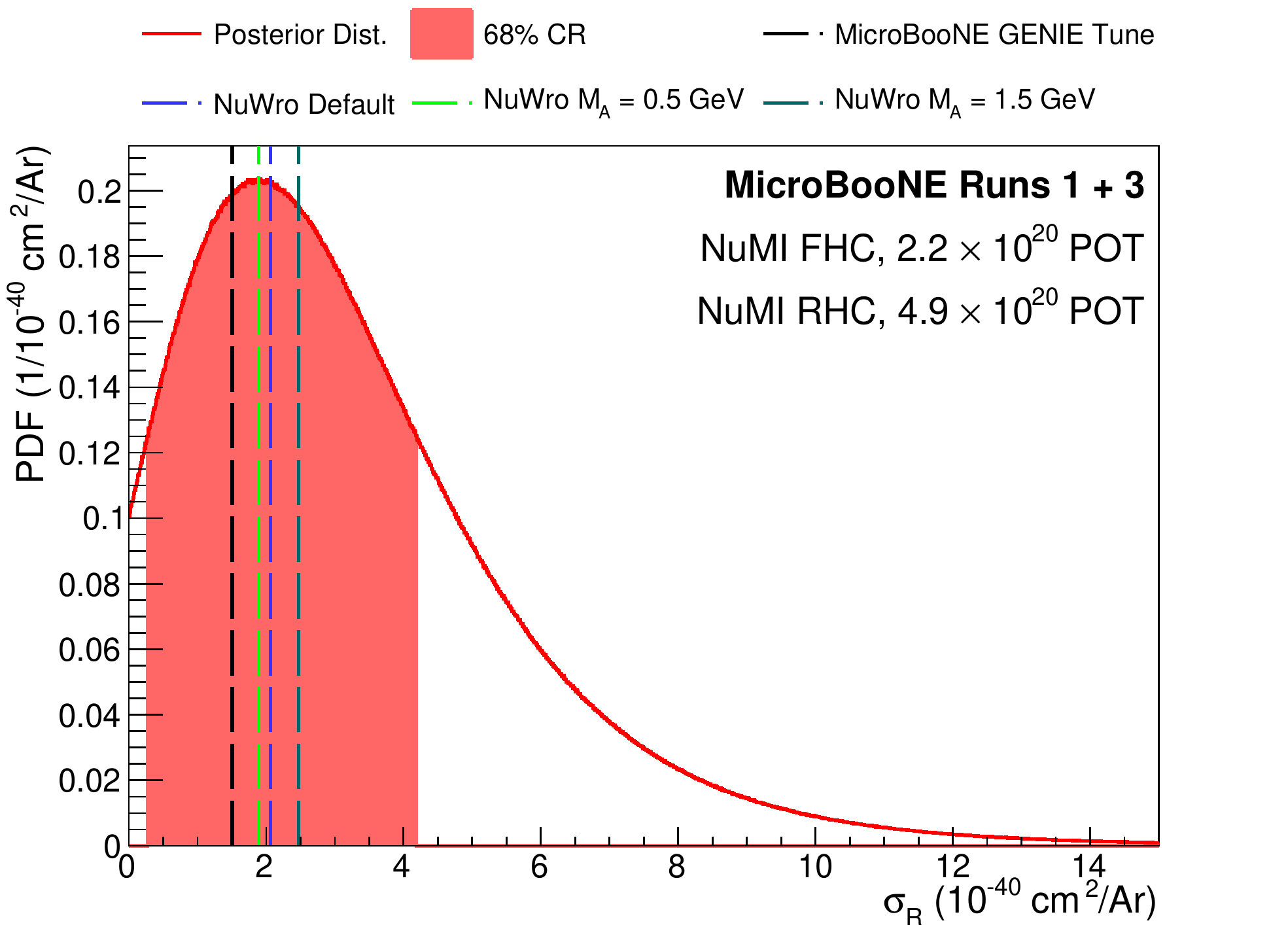}
    \caption{Posterior distribution describing the extracted cross section compared with the MicroBooNE GENIE tune~\cite{MicroBooNE:2021ccs} and three predictions from the NuWro event generator. The NuWro predictions include the effect of final state interactions, while GENIE does not take them into account. The standard axial mass used by NuWro is $1.03$ GeV.}
    \label{fig:PosteriorDist}
\end{figure*}

Fig.~\ref{fig:PosteriorDist} shows the Bayesian posterior probability distribution on the restricted phase space cross section, which is extracted by repeatedly generating values of $N$, $B$, and $\epsilon$ according to the Bayesian prior distributions obtained above. Each time these values are generated, a value of $\sigma_R$ is calculated according to
\begin{align} \label{eq}
\sigma_R = \frac{N - (B+B_0\alpha_{B})}{T(\Phi+\Phi_0\alpha_{\Phi})\Gamma(\epsilon+\epsilon_0\alpha_{\epsilon})}.
\end{align}
The fractional covariance matrix in Table~\ref{tab:CovarianceMatrix} is combined with the central values of $B$, $\Phi$, and $\epsilon$, to obtain their total covariance matrix, which is used to construct a three dimensional Gaussian distribution from which the systematic shifts, $\alpha_B$, $\alpha_{\Phi}$, and $\alpha_{\epsilon}$, applied respectively to $B$, $\Phi$, and $\epsilon$ are sampled. The uncertainties in $T$ and $\Gamma$ are assumed to be negligible.

\begin{figure}[h]
    \centering
    \includegraphics[width=\linewidth]{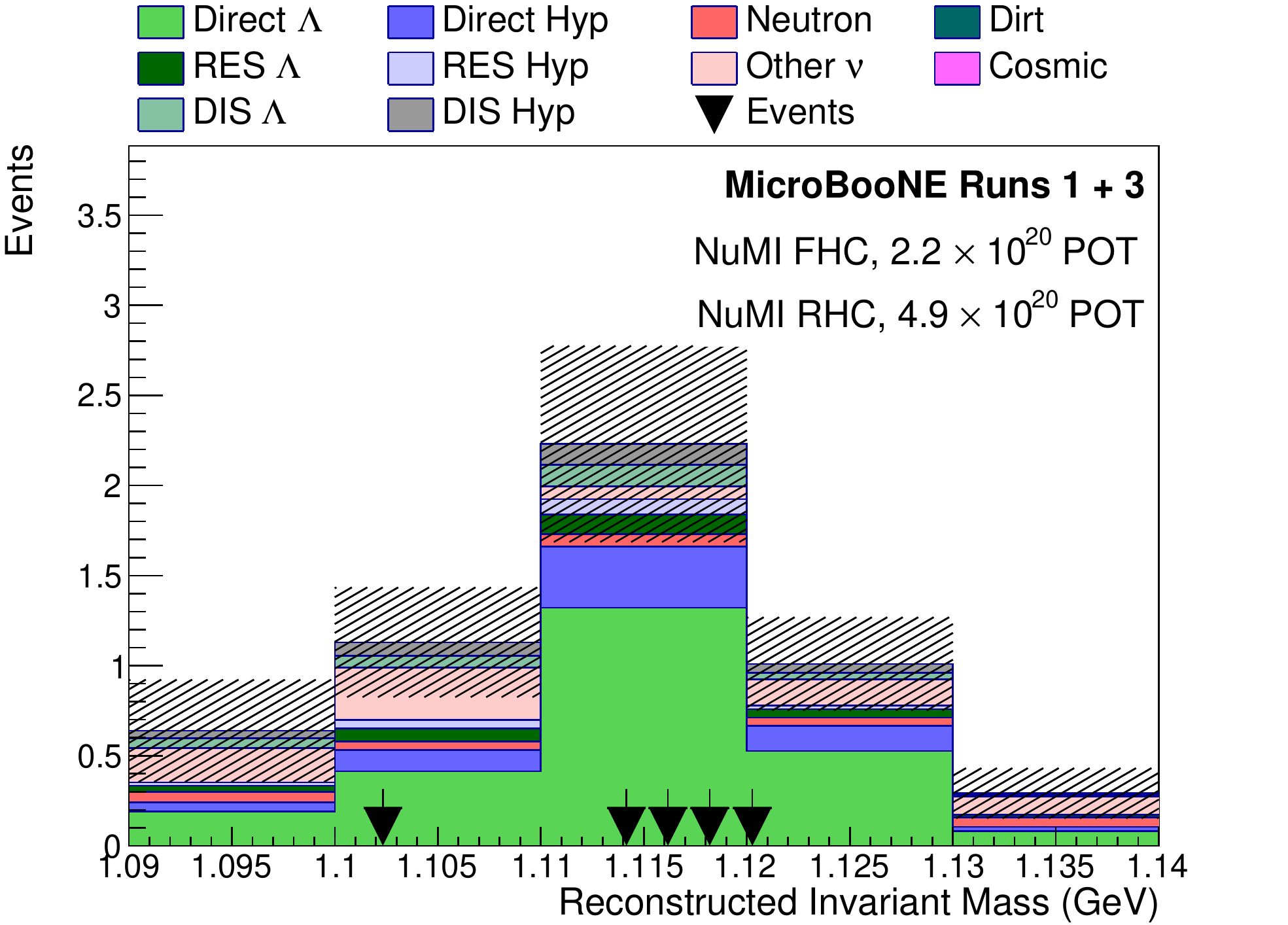}
    \caption{Selected MC simulation events and data, shown as a function of the reconstructed invariant mass, when using the purely automated selection. Black triangles indicate the locations of the selected data events. The mass of the $\Lambda$ baryon is 1.115 GeV~\cite{ParticleDataGroup:2020ssz}. The hatched regions indicate combined statistical and systematic uncertainties.}
    \label{fig:GENIE_BothRuns_InvariantMass}
\end{figure}

After unblinding the data in the signal region, 5 events are selected by the automated selection. The invariant masses of these events are compared with MC simulation predictions in Fig.~\ref{fig:GENIE_BothRuns_InvariantMass}. The five hand scanners selected 3, 3, 4, 4, and 5 events from the signal region. To extract the final cross section posterior distribution, we sum the Bayesian posterior distributions corresponding to observing those numbers of events and normalize the result to 1; the resulting distribution is shown in Fig.~\ref{fig:PosteriorDist}. The uncertainty in the cross section is obtained by constructing a 68\% credible interval from this distribution. We obtain a cross section of  $2.0^{+2.2}_{-1.7} \times 10^{-40}$ cm$^{2}/$Ar (combining statistical and systematic uncertainties), a value consistent with predictions from the GENIE~\cite{GENIE} and NuWro~\cite{Thorpe:2020tym} event generators. If we only apply the statistical fluctuations in Eq.~\ref{eq}, we obtain uncertainties of $^{+2.0}_{-1.4}$, while if only the systematic fluctuations are included, the uncertainties are $^{+1.2}_{-1.0}$, indicating the statistics are the dominant source of uncertainty.

In summary, the first measurement of the rare channel of quasi-elastic-like $\Lambda$ production in a LArTPC, using a mostly automated selection, has been performed. As this is a rare channel, the dominant source of uncertainty is data statistics. The adoption of a dedicated reconstruction algorithm for secondary vertices may lead to some improvement in the selection efficiency, but this will require significant development and is therefore beyond the scope of this work. Data collected between 2017 and 2020 awaits analysis, with which an approximately fourfold increase in signal events is expected.

This document was prepared by the MicroBooNE collaboration using the resources of the Fermi National Accelerator Laboratory (Fermilab), a U.S. Department of Energy, Office of Science, HEP User Facility. Fermilab is managed by Fermi Research Alliance, LLC (FRA), acting under Contract No. DE-AC02-07CH11359. MicroBooNE is supported by the following: the U.S. Department of Energy, Office of Science, Offices of High Energy Physics and Nuclear Physics; the U.S. National Science Foundation; the Swiss National Science Foundation; the Science and Technology Facilities Council (STFC), part of the United Kingdom Research and Innovation; the Royal Society (United Kingdom); and the UK Research and Innovation (UKRI) Future Leaders Fellowship. Additional support for  the laser calibration system and cosmic ray tagger was provided by the Albert Einstein Center for Fundamental Physics, Bern, Switzerland. We  also acknowledge the contributions of technical and scientific staff to the design, construction, and operation of the MicroBooNE detector as well as the contributions of past collaborators to the development of MicroBooNE analyses, without whom this work would not have been possible. For the purpose of open access, the authors have applied a Creative Commons Attribution (CC BY) public copyright license to any Author Accepted Manuscript version arising from this submission.


\begin{thebibliography}{10}

\bibitem{MicroBooNE:2016pwy}
R.~Acciarri \textit{et al.} (MicroBooNE),
Design and Construction of the MicroBooNE detector,
J. Instrum. \textbf{12}, P02017 (2017).

\bibitem{MINERvA:2016iqn}
L.~Aliaga \textit{et al.} (MINERvA),
Neutrino Flux Predictions for the NuMI Beam,
Phys. Rev. D \textbf{94}, 092005 (2016).

\bibitem{Adamson:2015dkw}
P.~Adamson \textit{et al.},
The NuMI Neutrino Beam,
Nucl. Instrum. Meth. A \textbf{806}, 279 (2016).

\bibitem{Ammosov:1986jn} 
V.~V.~Ammosov {\it et al.},
Neutral Strange Particle Exclusive Production in Charged Current High-energy Anti-neutrino Interactions,
Z.\ Phys.\ C {\bf 36}, 377 (1987).
  
\bibitem{Eichten:1972bb} 
T.~Eichten {\it et al.},
Observation of `Elastic' Hyperon Production by Anti-neutrinos,
Phys.\ Lett.\  {\bf 40} B, 593 (1972).
  
\bibitem{Erriquez1} 
O.~Erriquez {\it et al.},
Production of Strange Particles in Anti-neutrino Interactions at the CERN PS,
Nucl.\ Phys.\ B {\bf 140}  123 (1978).

\bibitem{Erriquez2} 
O. ~Erriquez {\it et al.},
Strange Particle Production by Antineutrinos,
Phys. Lett. B {\bf 70} 383 (1977).
  
\bibitem{Brunner:1989kw} 
J.~Brunner {\it et al.} (SKAT),
Quasielastic Nucleon and Hyperon Production by Neutrinos and Anti-neutrinos With Energies Below 30 GeV,
Z.\ Phys.\ C {\bf 45} 551 (1990).

\bibitem{Fanourakis:1980} 
G.~Fanourakis {\it et al.},
Study of Low-Energy Antineutrino Interactions on Protons,
Phys.\ Rev.\ D {\bf 21}, 562 (1980).

\bibitem{Thorpe:2020tym}
C.~Thorpe \textit{et al.},
Second Class Currents, Axial Mass, and Nuclear Effects in Hyperon Production,
Phys. Rev. C \textbf{104}, 035502 (2021).

\bibitem{Sobczyk:2019uej}
J.~E.~Sobczyk, N.~Rocco, A.~Lovato and J.~Nieves,
Weak Production of Strange and Charmed Ground-State Baryons in Nuclei,
Phys. Rev. C \textbf{99}, 065503 (2019).

\bibitem{Singh:2006xp}
S.~K.~Singh and M.~J.~Vicente Vacas,
Weak Quasi-elastic Production of Hyperons,
Phys. Rev. D \textbf{74}, 053009 (2006).

\bibitem{DUNE:2020ypp}
B.~Abi \textit{et al.} (DUNE),
Deep Underground Neutrino Experiment (DUNE), Far Detector Technical Design Report, Volume II: DUNE Physics,
arXiv:2002.03005 [hep-ex].

\bibitem{DUNE:2020fgq}
B.~Abi \textit{et al.} (DUNE),
Prospects for Beyond the Standard Model Physics Searches at the Deep Underground Neutrino Experiment,
Eur. Phys. J. C \textbf{81}, 322 (2021).

\bibitem{Hyper-Kamiokande:2018ofw}
K.~Abe \textit{et al.} (Hyper-Kamiokande),
Hyper-Kamiokande Design Report,
arXiv:1805.04163 [physics.ins-det].

\bibitem{MicroBooNE:2021zul}
P.~Abratenko \textit{et al.} (MicroBooNE),
Cosmic Ray Background Rejection with Wire-Cell LArTPC Event Reconstruction in the MicroBooNE Detector,
Phys. Rev. Applied \textbf{15}, 064071 (2021).

\bibitem{GENIE}
C.~Andrepoulos \textit{et al.},
The GENIE Neutrino Monte Carlo Generator,
Nucl. Instrum. Meth. A \textbf{614} 87 (2010).

\bibitem{GEANT4:2002zbu}
S.~Agostinelli \textit{et al.} (GEANT4),
{\sc Geant}~4--A Simulation Toolkit,
Nucl. Instrum. Meth. A \textbf{506}, 250 (2003).


\bibitem{MicroBooNE:2017xvs}
R.~Acciarri \textit{et al.} (MicroBooNE),
The Pandora Multi-algorithm Approach to Automated Pattern Recognition of Cosmic-ray Muon and Neutrino Events in the MicroBooNE Detector,
Eur. Phys. J. C \textbf{78}, 82 (2018).

\bibitem{MiniBooNE:2008hfu}
A.~A.~Aguilar-Arevalo \textit{et al.} (MiniBooNE),
The Neutrino Flux prediction at MiniBooNE,
Phys. Rev. D \textbf{79}, 072002 (2009).


\bibitem{MicroBooNE:2021ddy}
P.~Abratenko \textit{et al.} (MicroBooNE),
Calorimetric Classification of Track-like Signatures in Liquid Argon TPCs using MicroBooNE Data,
J. High Energy Physics \textbf{12}, 153 (2021).

\bibitem{Hocker:2007ht}
A.~Hoecker \textit{et al.},
TMVA - Toolkit for Multivariate Data Analysis,
arXiv:physics/0703039 [physics.data-an].

\bibitem{VanDePontseele:2020tqz}
W.~Van De Pontseele, Ph.D. thesis,
Search for Electron Neutrino Anomalies with the MicroBooNE Detector,
Oxford U (2020),
FERMILAB-THESIS-2020-11.

\bibitem{MicroBooNE:2018swd}
C.~Adams \textit{et al.} (MicroBooNE),
Ionization Electron Signal Processing in Single Phase LArTPCs. Part I. Algorithm Description and Quantitative Evaluation with MicroBooNE Simulation,
J. Instrum. \textbf{13}, P07006 (2018).


\bibitem{MicroBooNE:2018swe}
C.~Adams \textit{et al.} (MicroBooNE),
Ionization Electron Signal Processing in Single Phase LArTPCs. Part II. Data/simulation Comparison and Performance in MicroBooNE,
J. Instrum. \textbf{13}, P07007 (2018).

\bibitem{MicroBooNE:2017qiu}
R.~Acciarri \textit{et al.} (MicroBooNE),
Noise Characterization and Filtering in the MicroBooNE Liquid Argon TPC,
J. Instrum. \textbf{12}, P08003 (2017).

\bibitem{SupplementalMaterial}
P.~Abratenko \textit{et al.} (MicroBooNE),
Supplemental Material at [URL to be inserted by publisher]. 

\bibitem{AliagaSoplin:2016shs}
L.~Aliaga Soplin, Ph.D. thesis,
Neutrino Flux Prediction for the NuMI Beamline,
College of William and Mary (2016),
FERMILAB-THESIS-2016-03.

\bibitem{MicroBooNE:2021ccs}
P.~Abratenko \textit{et al.} (MicroBooNE),
New CC0$\pi$ GENIE Model Tune for MicroBooNE,
Phys. Rev. D \textbf{105}, 072001 (2022).

\bibitem{Calcutt:2021zck}
J.~Calcutt \textit{et al.},
Geant4Reweight: A Framework for Evaluating and Propagating Hadronic Interaction Uncertainties in Geant4,
J. Instrum. \textbf{16}, P08042 (2021).

\bibitem{CAPTAIN:2019fxo}
B.~Bhandari \textit{et al.} (CAPTAIN),
First Measurement of the Total Neutron Cross Section on Argon Between 100 and 800 MeV,
Phys. Rev. Lett. \textbf{123}, 042502 (2019).

\bibitem{MicroBooNE:2021roa}
P.~Abratenko \textit{et al.} (MicroBooNE),
Novel Approach for Evaluating Detector-Related Uncertainties in a LArTPC Using MicroBooNE Data,
Eur. Phys. J. C \textbf{82}, 454 (2022).

\bibitem{MicroBooNE:2019efx}
C.~Adams \textit{et al.} (MicroBooNE),
Calibration of the Charge and Energy Loss Per Unit Length of the MicroBooNE Liquid Argon Time Projection Chamber Using Muons and Protons,
J. Instrum. \textbf{15}, P03022 (2020).

\bibitem{MicroBooNE:2019koz}
C.~Adams \textit{et al.} (MicroBooNE),
A Method to Determine the Electric Field of Liquid Argon Time Projection Chambers Using a UV Laser System and its Application in MicroBooNE,
J. Instrum. \textbf{15}, P07010 (2020).

\bibitem{MicroBooNE:2020kca}
P.~Abratenko \textit{et al.} (MicroBooNE),
Measurement of Space Charge Effects in the MicroBooNE LArTPC Using Cosmic Muons,
J. Instrum. \textbf{15}, P12037 (2020).

\bibitem{ParticleDataGroup:2020ssz}
P.~A.~Zyla \textit{et al.} (Particle Data Group),
Review of \mbox{Particle} Physics,
Prog. Theor. Exp. Phys. \textbf{2020}, 083C01 (2020).

\bibitem{TEfficiency}
\url{https://root.cern.ch/doc/master/classTEfficiency.html}, accessed May 2022. Root version 6.16 used.

\end{thebibliography}
\end{document}